\documentclass[sigconf]{acmart}

\usepackage{tabulary}
\usepackage{multirow}
\usepackage{enumitem}
\usepackage{array}
\usepackage{caption}
\usepackage{subcaption}
\usepackage{makecell}
\usepackage{csquotes}
\AtBeginDocument{%
  }

\setcopyright{acmlicensed}
\copyrightyear{2024}
\acmYear{2024}
\acmDOI{XXXXXXX.XXXXXXX}

\acmConference[Conference LAK '25]{Make sure to enter the correct
  conference title from your rights confirmation email}{June 03--05,
  2018}{Woodstock, NY}
\acmISBN{978-1-4503-XXXX-X/18/06}

\begin{document}

\title{Designing Prompt Analytics Dashboards to Analyze Student-ChatGPT Interactions in EFL Writing}
\author{Minsun Kim}
\affiliation{%
  \institution{Industrial Design, KAIST}
  \country{Daejeon, South Korea}
}
\email{9909cindy@kaist.ac.kr} 

\author{SeonGyeom Kim}
\affiliation{%
  \institution{Industrial Design, KAIST}
  \country{Daejeon, South Korea}
}
\email{sgk_0320@kaist.ac.kr}

\author{Suyoun Lee}
\affiliation{%
  \institution{Industrial Design, KAIST}
  \country{Daejeon, South Korea}
}
\email{jenslee705@kaist.ac.kr}

\author{Yoosang Yoon}
\affiliation{%
  \institution{Industrial Design, KAIST}
  \country{Daejeon, South Korea}
}
\email{soulmilk98@kaist.ac.kr}

\author{Junho Myung}
\affiliation{%
  \institution{School of Computing, KAIST}
  \country{Daejeon, South Korea}
}
\email{junho00211@kaist.ac.kr}

\author{Haneul Yoo}
\affiliation{%
  \institution{School of Computing, KAIST}
  \country{Daejeon, South Korea}
}
\email{haneul.yoo@kaist.ac.kr}

\author{Hyunseung Lim}
\affiliation{%
  \institution{Industrial Design, KAIST}
  \country{Daejeon, South Korea}
}
\email{charlie9807@kaist.ac.kr}

\author{Jieun Han}
\affiliation{%
  \institution{School of Computing, KAIST}
  \country{Daejeon, South Korea}
}
\email{jieun_han@kaist.ac.kr}

\author{Yoonsu Kim} 
\affiliation{%
  \institution{School of Computing, KAIST}
  \country{Daejeon, South Korea}
}
\email{yoonsu16@kaist.ac.kr}

\author{So-Yeon Ahn}
\affiliation{%
  \institution{School of Digital Humanities and Computational Social Sciences, KAIST}
  \country{Daejeon, South Korea}
}
\email{ahnsoyeon@kaist.ac.kr}

\author{Juho Kim}
\affiliation{%
  \institution{School of Computing, KAIST}
  \country{Daejeon, South Korea}
}
\email{juhokim@kaist.ac.kr}

\author{Alice Oh}
\affiliation{%
  \institution{School of Computing, KAIST}
  \country{Daejeon, South Korea}
}
\email{alice.oh@kaist.edu}
\author{Hwajung Hong}
\affiliation{%
  \institution{Industrial Design, KAIST}
  \country{Daejeon, South Korea}
}
\email{hwajung@kaist.ac.kr}
\author{Tak Yeon Lee}
\affiliation{%
  \institution{Industrial Design, KAIST}
  \country{Daejeon, South Korea}
}
\email{takyeonlee@kaist.ac.kr}
\renewcommand{\shortauthors}{Kim et al.}

\begin{abstract}
While ChatGPT has significantly impacted education by offering personalized resources for students, its integration into educational settings poses unprecedented risks, such as inaccuracies and biases in AI-generated content, plagiarism and over-reliance on AI, and privacy and security issues. To help teachers address such risks, we conducted a two-step design process that comprises interviews, and a prototype demonstration involving six EFL (English as a Foreign Language) teachers, who integrated ChatGPT into semester-long English essay writing classes. Based on the design goals identified during the initial interviews, we developed a prototype of Prompt Analytics Dashboard (PAD) that integrates the essay editing history and chat logs between students and ChatGPT. 
Teacher's feedback on the prototype informs additional features and unmet needs for designing future PAD, which helps them (1) analyze contextual analysis of student behaviors, (2) design an overall learning loop, and (3) develop their teaching skills.
\end{abstract}



\received{20 February 2007}
\received[revised]{12 March 2009}
\received[accepted]{5 June 2009}

\maketitle

\section{Introduction}
ChatGPT emerges as a transformative tool in education \cite{kasneci_chatgpt_2023}, where a fast-growing number of students and teachers adopted ChatGPT to generate personalized feedback and assist in the creation of learning materials at any time and place \cite{fuchs_exploring_2023, baidoo-anu_education_2023}. In particular, ChatGPT has been successful in the domain of language learning, where students with different levels of language proficiency can benefit from customized resources \cite{baskara2023exploring}.

While ChatGPT offers exciting opportunities, there exist a variety of challenges to using it in the field of education. Along with other LLM-based services, ChatGPT is prone to produce inaccurate information, hallucinations, and a wide range of biases \cite{gupta_chatgpt_2023, alkaissi_artificial_2023, ray_chatgpt_2023, rahman_chatgpt_2023}. Educational experts also have warned that an over-reliance on ChatGPT may undermine students' critical thinking and problem-solving abilities \cite{rahman_chatgpt_2023, ray_chatgpt_2023}. In particular, students who blindly copy and paste the answers from ChatGPT will miss the learning opportunity.

To analyze student behaviors and to reflect on their practices, teachers have been employing Learning Analytics Dashboards (LAD) even before ChatGPT was invented \cite{schwendimann_perceiving_2017}. With an extensive amount of chat logs between students and ChatGPT, we expect that the role of analytic tools will only grow bigger \cite{han_recipe_2023}. Nevertheless, there is a significant knowledge gap between the design of conventional LAD and future classrooms integrated with ChatGPT. For example, it is unclear how an analytic dashboard can help teachers gain an in-depth understanding of student behavior, prevent students' over-reliance and misuse, and facilitate effective use of ChatGPT. The role of a dashboard in effectively integrating ChatGPT into future educational paradigms is yet to be determined. 

In this paper, we introduce Prompt Analytics Dashboard (PAD), an interactive tool for teachers to track, analyze, and respond to student-ChatGPT interaction. To examine its roles and design implications grounded in real teaching experiences rather than hypothetical scenarios, we conducted a design process alongside semester-long English for Foreign Languages (EFL) writing courses.

The contribution of this paper is three-fold. Firstly, we investigated teachers' perceptions of the challenges and impacts of using ChatGPT in the context of language learning. This process helped us identify the expected benefits, required features, and design implications for the Prompt Analytics Dashboard (PAD). Secondly, we developed the PAD prototype, which supports teachers in analyzing contextual factors behind student behaviors, designing an overall learning loop, and enhancing their teaching practices. Finally, we envisioned how the integration of PAD can promote more effective teaching strategies and better support student learning outcomes in future EFL classes.

\section{Related Work}
\subsection{Large Language Model Systems in Language Education}
In recent years, Large Language Models (LLMs), exemplified by the Generative pretrained Transformer (GPT-3) \cite{floridi_gpt-3_2020}, have achieved significant progress in natural language processing (NLP). At this moment, ChatGPT, an LLM-driven chatbot, is capable of understanding and responding to almost natural human language \cite{pradana_discussing_2023}, and it could support various tasks related to text-based inputs.

In the context of language education, the empirical studies on leveraging ChatGPT to develop a second language (L2) mostly focused on writing skills, vocabulary, and assessments \cite{yan_impact_2023, kohnke_chatgpt_2023-1}. ChatGPT can serve as a personal tutor for students, thereby answering their questions in real-time \cite{sabzalieva_2023_chatgpt}. RECIPE, a ChatGPT-integrated learning platform, is tailored for students in the context of EFL writing education \cite{han2023recipe}. From an instructor's perspective, ChatGPT improves the teaching experience by generating teaching and assessment methods for instructors, thereby shifting the paradigm of educators' role \cite{rudolph_chatgpt_2023}. 
For instance, LLMs can assist in developing teaching materials such as quizzes and assignments \cite{lu_2023_readingquizmaker}.

Despite these advantages, instructors face challenges in detecting plagiarism when students use ChatGPT, as the tool can seamlessly generate text that may not be easily distinguishable from student's writing \cite{cotton_2023_chatting}. This is particularly problematic for students who rely heavily on external sources for information acquisition, as it may impact their learning and understanding \cite{ahmad_2023_hallucinations}. \cite{rane_2023_chatgpt} therefore emphasizes the necessity of an effective mechanism for instructors to detect and monitor students' usage of LLM. Given these opportunities and challenges, it is crucial to understand instructors' perceptions and develop systems that alleviate their concerns regarding the use of ChatGPT in educational settings.

\subsection{Human-Centered Learning Analytics}

Despite the advancement of learning technology leading to the development of many LADs, only a few are actually utilized in real-world settings.
To mitigate this gap, the concept of `human-centered learning analytics' (HCLA) was recently introduced, emphasizing the consideration of human aspects in the design of LADs \cite{ahn_designing_2019}. The design methods for the pre-existing dashboard were focused on enhancing usability and usefulness \cite{verbert_learning_2013}, reducing time spent on data skimming, and aiding in the accurate interpretation of information \cite{kaliisa_cada_2023}. Compared to previous dashboard design methods, the HCLA dashboard keeps an active involvement of educational stakeholders in the design process and the incorporation of their suggestions into the visual outputs \cite{alfredo2024slade}. These approaches offer the advantage of differing from traditional technically inclined methods, better reflecting the perspectives of actual users.

Recent studies have applied HCLA in three distinct approaches \cite{pozdniakov_question-driven_2021}: co-design with stakeholders to understand their authentic needs, employing theoretical and methodological approaches to enhance the HCLA process, and focusing on data literacy in the context of human perception. The background of this approach is from human-computer interaction, focusing on human values and concerns at the center of technology design and assessment, with co-design methods integrating user participation \cite{sanders_co-creation_2008}.
In this context, HCLA must consider designing with educational stakeholders who lack expertise in data and analytics \cite{pozdniakov_question-driven_2021}.

In this paper, we focus on designing human-centered PAD by adopting HCLA with an iterative design process but primarily focus on qualitatively finding implications through a dashboard prototype by adopting a real-world class environment dataset.

\section{Method}
\subsection{A Semester-long Essay Writing Course with ChatGPT}
The study aims to get insights for designing PAD (Prompt Analytic Dashboard), which is grounded in real teaching experiences rather than hypothetical scenarios. Therefore the design process was performed alongside semester-long English for Foreign Languages (EFL) writing courses. During the courses, students were required to write and revise their essays based on weekly learning objectives.

For seamless integration of ChatGPT with the essay writing process, students used EWP\footnote{The actual name and the citation of the tool are removed for the double-blind review process.}, a web-based essay writing platform, where students can use ChatGPT API while composing and revising their essays throughout the semester. We refer to this repetitive writing process as ``ChatGPT-student interaction.''  Instead of using the off-the-shelf ChatGPT, the platform aligns the API's behavior to support English writing education and to lead the students, by using hidden base prompts. For example, at the beginning of each session, it proactively prompts students to review the learning objectives instead of passively waiting for student requests. 

\subsection{Participants}
We recruited six university English teachers (see Table \ref{tab:demograhics} for demographics), who were teaching the aforementioned essay writing courses alongside the study. They had an average of 17 years of teaching experience (MIN = 6, MAX = 26). The courses were categorized based on students' English proficiency levels: intermediate and advanced writing classes for undergraduates, and an academic writing class for graduate students. All teachers had taught English courses across various proficiency levels. Two of them held degrees in STEM fields, while the others specialized in teaching English. Regarding statistical knowledge, two teachers lacked a background in statistics, whereas the others had some familiarity through their academic experiences. All teachers consented to participate in this study, which received approval from our institutional review board.

\begin{table}
\centering
\caption{Participant demographics}
\label{tab:demograhics}
\resizebox{\columnwidth}{!}{%
\begin{tabular}{c|c|c|c|c|l}
\toprule
\textbf{Participant ID} &
  \textbf{Age Group} &
  \textbf{Gender} &
  \textbf{\makecell{Years in \\ Teaching}} &
  \textbf{\makecell{STEM}} &
  \textbf{\makecell{Statistical background}} \\ \midrule
    P1 &
    30-39 &
    Female &
    6 &
    Non-STEM&
    No statistical background \\ 
    P2 & 
    40-49 & 
    Male   & 
    25 & 
    Non-STEM&
    Statistical experience via research \\ 
    P3 & 
    40-49 & 
    Female & 
    15 & 
    STEM&
    Statistical experience on big-data \\ 
    P4 &
    40-49 &
    Female &
    17 &
    STEM&
    Statistical background on her major \\ 
    P5 &
    40-49 &
    Male   &
    13 &
    Non-STEM&
    Statistical experience via research \\ 
    P6 &
    50-59 &
    Male   &
    26 &
    Non-STEM&
    No statistical background\\ \bottomrule
\end{tabular}%
}
\end{table}

\subsection{Design Process}
To collect insights based on actual teaching experiences, we divided the study into two steps, corresponding to the semester-long course timeline.

The first step of the study was conducted two months after the semester began to wait until teachers accumulated enough experience in applying ChatGPT (via EWP). It aimed for two types of findings: (1) teachers' perception of using ChatGPT in English education, and (2) initial design inspirations of PAD by observing how teachers track, evaluate, and guide student-ChatGPT interaction. This study was conducted in 90 minutes and comprises three steps, as reported in Section \ref{section:interview_process1}. 

Based on the insights gathered, we developed a web-based prototype of PAD (Prompt Analytics Dashboard) for teachers, before the second step of the study. The prototype features a wide range of charts and other components, displaying actual essays and log data collected along the semester, as detailed in Section \ref{section:dashboard_prototype}.

To collect feedback on the prototype and explore the potential impact of PAD in future EFL classes, we initiated the second step of the study right after the semester ended. In this step, we introduced the prototype, allowing teachers to try out the dashboard's components. Finally, we conducted a semi-structured interview, as detailed in Section \ref{section:interview_process2}. 

During the entire study, we collected voice recordings of the interviews. While transcribing the interviews, we also made notes based on our observations. After conducting a qualitative analysis using this dataset, we derived design implications by synthesizing the outputs from both steps. For detailed information on methods and analysis procedures, please refer to the respective sections for each step.

\section{Needs Finding(Interview 1)}
\subsection{Interview Process}\label{section:interview_process1}
\subsubsection{Examining teachers' perspective}
To begin with, we carried out a brief survey and follow-up interviews to explore teachers' viewpoints and concerns about using ChatGPT in their classrooms. This survey was used as a means to spark discussion about ChatGPT rather than to extract statistical data. During the survey, teachers evaluated ChatGPT's utility in fostering a variety of English skills - such as reading, grammar, vocabulary, writing, listening, and speaking \cite{hasbullah_developing_2023} on a 7-point Likert scale that varied from -3(Very Unhelpful) to 3(Very Helpful). Subsequently, they evaluated a variety of potential risks associated with allowing students to use ChatGPT for their essay writing exercises - such as plagiarism, biasing, lack of information, lack of empathy, inconsistency, ChatGPT misinterpreting, ChatGPT may leak sensitive student information, security concerns, no verification of answer quality, degrading critical thinking ability, and inaccurate answer - as identified in \cite{rane_2023_chatgpt}.
After teachers completed the survey, we conducted follow-up interviews that inquired what specific experience affected their choices, and how their students had been using the EWP platform so far.

\subsubsection{Observing how teachers evaluate student-ChatGPT interaction}
\begin{figure}
     \centering
     \begin{subfigure}[b]{0.49\textwidth}
         \centering
         \includegraphics[width=0.9\textwidth]{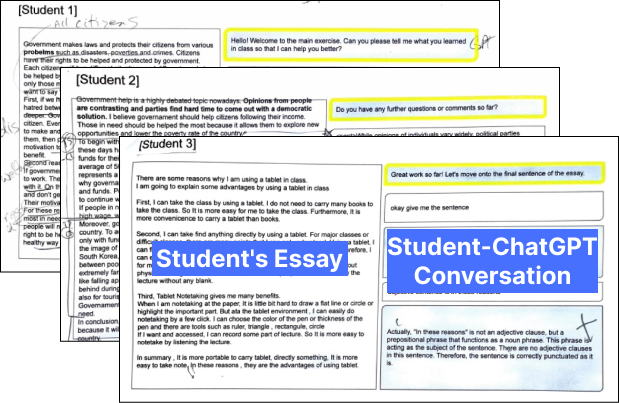}
         \caption{Examples of student-ChatGPT interaction.}
         \label{fig:handout_explanation}
     \end{subfigure}
    \begin{subfigure}[b]{0.49\textwidth}
        \centering
        \includegraphics[width=0.9\textwidth]{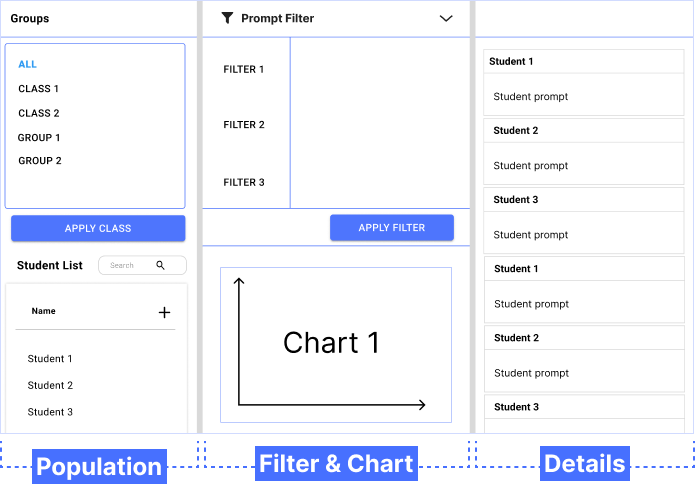}
        \caption{Three-steps dashboard sketch.}
        \label{fig:3-step_sketch}
    \end{subfigure}
    \caption{Contents of interview handouts.}
    \Description{Contents of interview handouts.}
    \label{fig:handouts}
\end{figure}

The observation session aims to gather insights that would inform the design objectives of PAD. To achieve this goal, we observed how teachers would evaluate students' essay editing history along with corresponding chat logs. Before the observation, we selected three randomly chosen students and prepared hard copies of their chat logs and the corresponding essays.

These hard copies were designed with a specific layout: on the left side, the essay logs were presented, and on the right side, the chat logs were displayed. To clearly illustrate the correlation between chat log changes and essay edits, we highlighted the chat boxes with a yellow marker in chronological order and bolded the points of change in the essays to denote significant edits. Additionally, the handout depicted the "edit history" not just by showing the essay and the prompts used, but by emphasizing the interactive changes over time. The highlighted chat boxes indicated moments where teacher prompts directly influenced subsequent edits in the student essays. This allowed teachers to easily trace and assess the impact of their feedback on student work.

To ensure that the selected students were representative of the class population, we validated that the chat logs contained distinct prompting patterns based on \cite{han_exploring_nodate} and that the quality of the essays varied widely, encompassing a broad spectrum of proficiency levels, from basic to advanced. During the review process, teachers were asked to highlight interesting sections and write memos on hard copies, simulating their routine of evaluating homework. They were also encouraged to think aloud, explaining why they considered certain behaviors educationally effective or not, and how they would respond to these behaviors.

\subsubsection{Gathering design needs}
Following the observation, we presented the rough initial sketch of the PAD's UI (Figure \ref{fig:3-step_sketch}), which is designed to support three typical stages of visual analytics: overview, zoom and filter, and detail-on-demand \cite{shneiderman_eyes_1996}. This initial sketch serves as a prototype to gather feedback from teachers and includes three primary components: overview charts, filters, and detailed information.

\subsection{Analysis}
With the interview transcripts and observational notes, we conducted a qualitative analysis with a mixed approach. To elaborate on the process, two researchers collaboratively analyzed the dataset, engaging in multiple rounds of open coding until they reached a consensus. First, they identified common concerns and challenges based on the interview transcripts. Subsequently, they extracted teachers' needs for PAD from the notes taken during the observational study. 

\subsection{Challenges and Needs on Analyzing ChatGPT-Student Interaction}

In this section, we present three significant challenges that teachers encountered in (1) assessing the effectiveness of learning, (2) handling excessive additional workload and responsibilities, and (3) addressing reliability, privacy, and security concerns. These challenges provided us with motivations for designing the prototype, presented in Section \ref{section:dashboard_prototype}. 

\subsubsection{Assessing the effectiveness of learning}
All the teachers involved in the study acknowledged ChatGPT's proficiency in a wide range of text generation tasks, such as producing phrases for learning new vocabulary, rephrasing or translating text, or generating paragraphs for given topics. Nevertheless, they were disappointed by ChatGPT's incompetence in accounting for educational outcomes. As highlighted by five teachers, ChatGPT encourages students to use ideas offered by external sources rather than to formulate their own opinions, thereby diminishing students' opportunities to think critically.  
In particular, all the teachers were seriously concerned about students' behavior of asking ChatGPT to write paragraphs or even an entire essay, ``I observed a few students copy-and-pasted AI-generated paragraphs without reading them. That was not just plagiarism but also harmful to their critical thinking,'' as P1 said. Hence, a majority of teachers advised students against requesting ChatGPT to generate paragraphs, flagging such behaviors as non-educational. 

Assessing a prompt's educational effectiveness may not be straightforward, given that the teacher must consider both the student's intent and the impact on learning. For example, according to P4's observation, students with lower writing proficiency frequently stuck at the beginning of writing their essays. Thus generated content would be rather beneficial if students used it as scaffolding of their essay structure. Similarly, P5 noted that translating text into English may hold educational value, particularly in certain situations, such as preparing speech scripts. In consequence, to precisely assess the educational efficacy of Student-ChatGPT interaction, teachers need to review the prompt, ChatGPT's response, and the corresponding segment of the essay editing history.

The last aspect that teachers cared about was the alignment between prompts and the learning objectives. Notably, every teacher emphasized that their classes focus on enhancing students' ability to construct essays with a clear and effective structure, rather than on learning advanced vocabulary or grammar. From that perspective, prompts about structural aspects of the essay must have greater potential values, compared to prompts about general knowledge.    
For example, P3 highlighted the need for filtering prompts with the learning objectives of the week, by saying, ``Using each chapter's keywords as filtering criteria, teachers should be able to see what students have learned this week, and how they incorporated it into their essays.''

To sum up, our findings suggest that the assessment of learning effectiveness requires a holistic examination of the prompts that students used, their application of the responses, and how these align with the learning objectives.

\subsubsection{Additional workload and responsibilities}\label{sec:workload}
Following the two months of integrating ChatGPT into their teaching practices, teachers felt an increased amount of extra work and responsibilities.
During the observation process, teachers said the most tedious and time-consuming task at hand was plagiarism detection. There exist even harder cases of subtle and nuanced plagiarism in which students made small modifications to the generated content. P6 pointed out that checking every potential case of plagiarism is almost infeasible, unless the copied text contains low-quality expressions obviously due to the technical limitations of ChatGPT, as P2 noted. 

Teachers anticipated an additional workload in instructing ChatGPT on the learning objectives of their classes. They also mentioned the lack of such contextual understanding is the root cause of ChatGPT misinterpreting students' inquiries, and thus generating irrelevant and non-educational responses. For example, P4 highlighted, ``ChatGPT is not good at aligning its feedback (on students' essays) to the lecture. Feedback is often too generic, and occasionally conflicting with the point that I taught.'' She also suggested that a more effective approach to instructing ChatGPT could involve uploading all lecture notes, which would enable ChatGPT to accurately interpret lecture-related questions, and generate responses aligned with the learning objectives.

Teaching the usage of ChatGPT during their lecture hours was another extra duty for teachers. In particular, they considered it essential to give students instructions, with positive and negative examples, on how to articulate one's intent through prompts and judge the quality of ChatGPT's responses. According to P3, even students with high proficiency in English may struggle with the above tasks. 

Lastly, evaluating homework involves extra tasks for teachers to review the edit history and corresponding chat log. Therefore, while acknowledging the benefits of the contextual information, all teachers were interested in how PAD will be able to enhance the efficiency of the homework evaluation process. For instance, a few teachers envisioned the system to automatically find essays having an identical issue (e.g., plagiarism), and generate feedback messages.     

\subsubsection{Promoting ethical and responsible utilization of ChatGPT}  
Given that the study was conducted alongside the college EFL program, fostering ethical and responsible utilization of ChatGPT within the academic context was a primary concern of the teachers. In particular, three teachers emphasized the importance of safeguarding sensitive or confidential information. For example, P3 is particularly worried that students' original research ideas in their prompts might become part of the model and eventually be shared with other ChatGPT users. P6 expressed similar concerns about the privacy and security issues, by saying, ``I don't know how private that is or if there are any security concerns.'' 

Biases in ChatGPT-generated content were another major concern, as it was trained on unvetted data from the internet, which might reflect a wide range of biases and stereotypes that people have. For instance, P3 pointed out that ChatGPT often generates politically biased responses, having potentially negative impacts on students. 

While teachers expressed concerns about inaccurate and unreliable information in responses generated by ChatGPT, they also pointed out internet search engines had the same issue - and it is individual students' responsibility to cross-validate any information obtained from external resources with textbooks, academic journals, or more reputable websites. Nonetheless, P5 suggested that the proliferation of AI-generated content soon may exacerbate the challenge of discerning reliable human-created content.

\section{Dashboard Prototype}
\subsection{Design Goals}
In developing the PAD, we aimed to address specific needs and challenges associated with integrating ChatGPT into educational settings. Each design goal is outlined with associated challenges, derived needs, and implemented features (Table \ref{tab:design-goal}).
\begin{table*}
\caption{Design goals.}
\label{tab:design-goal}
\resizebox{\linewidth}{!}{%
\begin{tabular}{c|c|c|c}
\toprule
\textbf{Design Goal} & \textbf{Challenges} & \textbf{Needs} & \textbf{Features} \\ \midrule
\textbf{\begin{tabular}[c]{@{}c@{}}Providing a quick overview\\ of student-ChatGPT interaction.\end{tabular}} & \begin{tabular}[c]{@{}c@{}}Teachers need an efficient \\ way to monitor multiple \\ students' interactions without \\ being overwhelmed by data.\end{tabular} & \begin{tabular}[c]{@{}c@{}}Tools to visualize chat frequency, \\ essay evaluations, \\ and misuse instances clearly and concisely.\end{tabular} & \begin{tabular}[c]{@{}c@{}}- Weekly chat frequency charts\\ - Essay evaluation summaries\\ - Misuse frequency charts\\ - 12 common patterns of \\ Student-ChatGPT dialogue \cite{han_exploring_nodate}\end{tabular} \\ \hline
\textbf{\begin{tabular}[c]{@{}c@{}}Identifying common patterns\\ in undesirable usage of ChatGPT.\end{tabular}} & \begin{tabular}[c]{@{}c@{}}Preventing students from \\ misusing ChatGPT \\ for non-educational purposes.\end{tabular} & \begin{tabular}[c]{@{}c@{}}Effective identification \\ and reporting of \\ undesirable usage patterns.\end{tabular} & \begin{tabular}[c]{@{}c@{}}- Highlighting and tagging of \\ non-educational prompts\\ - Automated detection of misuse patterns\end{tabular} \\ \hline
\textbf{\begin{tabular}[c]{@{}c@{}}Facilitating in-depth analysis \\ of essay editing and chat log.\end{tabular}} & \begin{tabular}[c]{@{}c@{}}Teachers need comprehensive insights \\ into student behavior to tailor \\ their instructional strategies.\end{tabular} & \begin{tabular}[c]{@{}c@{}}Integrated views of editing history \\ and chat logs, \\ ability to filter by learning objectives.\end{tabular} & \begin{tabular}[c]{@{}c@{}}- Displaying essay editing history \\ and corresponding chat logs\\ - Listing learning objectives for each class unit\\ - Filtering prompts by specific objectives\end{tabular} \\ \hline
\textbf{Customizing Prompt Instructions.} & \begin{tabular}[c]{@{}c@{}}Ensuring that ChatGPT provides \\ responses aligned with the \\ educational goals and learning objectives.\end{tabular} & \begin{tabular}[c]{@{}c@{}}Mechanisms to customize \\ ChatGPT instructions \\ based on teacher input.\end{tabular} & \begin{tabular}[c]{@{}c@{}}- Feature to add custom instructions that are \\ transmitted to the student-side system\end{tabular} \\ \bottomrule
\end{tabular}%
}
\end{table*}
\begin{figure*}
    \centering
    \includegraphics[width=\textwidth]{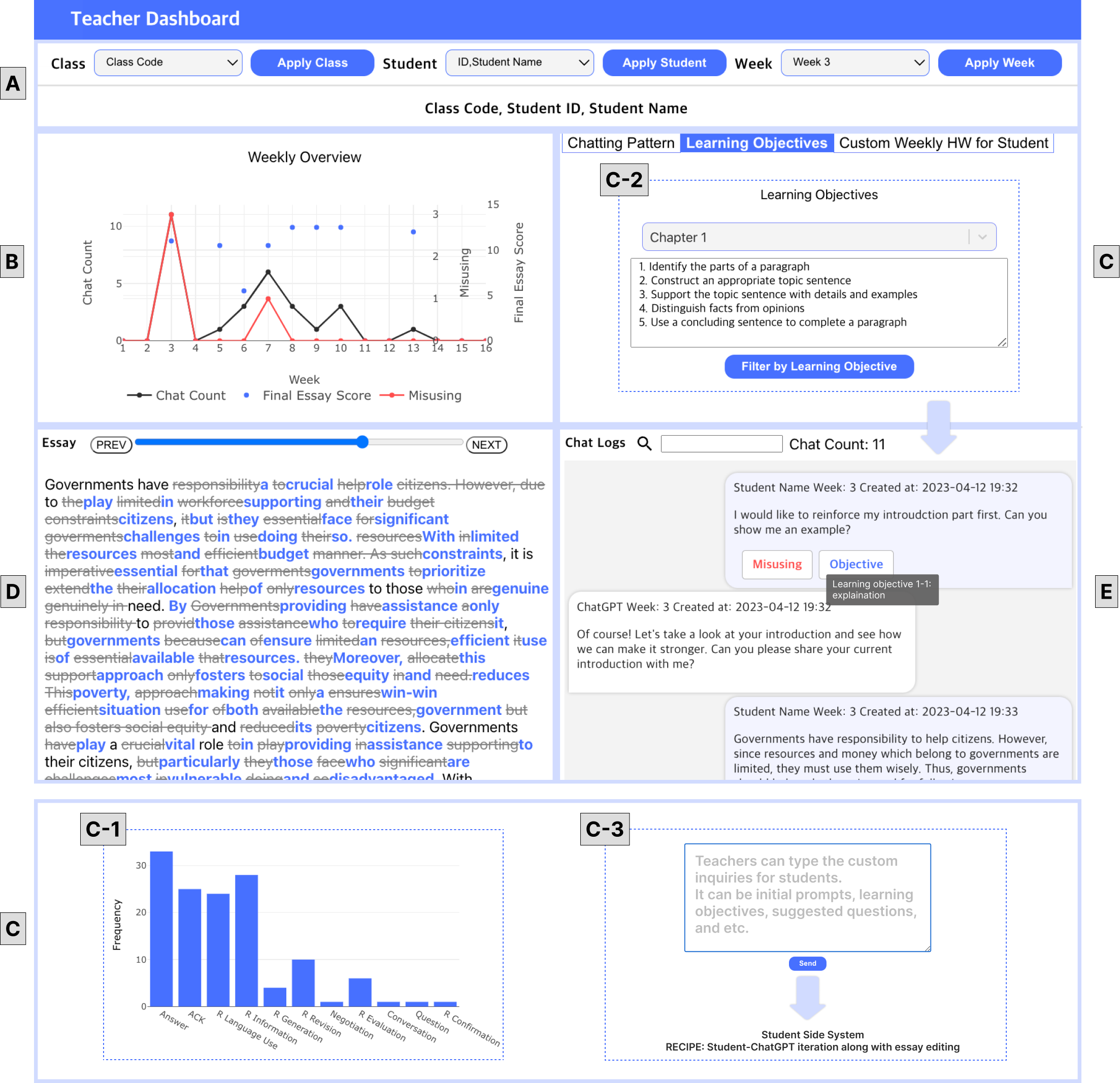}
    \caption{Screenshot of dashboard prototype. (a) Population pool: The dashboard displays information calculated based on the selected population pool. Teachers can narrow the analysis pool by selecting classes, students, and weeks. (b) Overview: This is an overview chart with the x-axis fixed to weekly, allowing you to see the chat count over time, the last saved essay score graded by AI, and the misuse count. (c) This area consists of three tabs, (c-1) a chart for chatting patterns, (c-2) additional filtering options based on learning objectives, and (c-3) the ability to deliver messages to students. (d) Teachers can track changes to essays by moving sliders. (e) Shows the students' entire chat history. Teachers can search, and each prompt has a tag so teachers can check the chat content.}
    \Description{Screenshot of dashboard prototype.}
    \label{fig:dashboard}
\end{figure*}
Based on the design goal, we developed a prototype of a teacher's dashboard (Figure \ref{fig:dashboard}), which supports them in analyzing essay writing and chat logs. 

\subsection{Dashboard Architecture}
\label{section:dashboard_prototype}

\begin{figure*}[t]
\centering
  \includegraphics[width=0.9\linewidth]{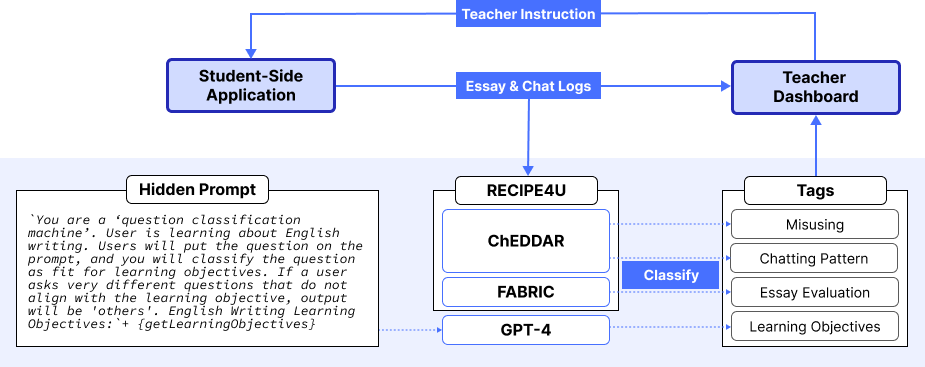} \hfill
  \caption {A dashboard system architecture.}
  \Description{The dashboard system architecture encompasses EWP, the student-side system collecting essays and chat logs, which are then analyzed by three AI models (RECIPE4U \cite{han2024recipe4u}, FABRIC \cite{han_2023_fabric}, GPT-4 \cite{OpenAI2023GPT4}). The teacher's dashboard integrates the tags and the raw data into various analytic components. }
  \label{fig:architecture}
\end{figure*}
The dashboard is part of a larger system, as outlined in Figure \ref{fig:architecture}, where students use a student-side application(EWP) to write essays every week and interact with ChatGPT. The essays and chat logs are automatically analyzed with NLP models.
To detect and highlight AI misuse and chatting patterns, we employed the ChEDDAR \citep{han_2023_cheddar}. This model is fine-tuned GPT-4 and identifies misuse by analyzing patterns of interaction between students and ChatGPT. Specifically, we defined misuse based on the most frequently occurring patterns among the twelve identified by the model. These patterns include requests for entire essay generation, excessive use of paraphrasing tools, and non-academic inquiries.
FABRIC contains an automated essay-scoring(AES) model based on teachers' essay grading data \citep{han2023fabric}. This is rubric-based AES using BERT \citep{devlin2018bert}. ChEDDAR and FABRIC results are combined into the RECIPE4U dataset, and utilized in the overall teacher dashboard. Also, \ref{fig:architecture} illustrates GPT-4 categorizes student prompts into learning objectives. This process is simply conducted with prompt engineering without evaluation. Learning objectives are from the university EFL lecture book.

\subsection{User Scenario}
To demonstrate the capability of the dashboard, this section presents the hypothetical scenario of Maria, a university EFL teacher, using the dashboard for her essay writing classes. Her students had been actively using EWP, the student-side essay writing platform integrated with ChatGPT API, for revising their essays.

\subsubsection{Investigation of students' behavior, from the overview to the chat log.}
During the third week of the semester, Maria grew interested in how students had been writing essays. She accessed the dashboard, reviewed the overview chart (B in Figure \ref{fig:dashboard}), and noticed a notable surge in prompts for Week 3, which suggests a favorable development. However, a further examination of the chart revealed that a lot of prompts used in Week 3 were highlighted in red, indicating that the AI models had flagged them as potential misuses. Maria checked the chatting patterns component (C-1) and figured out that most of them were flagged as ``Requests for Generation'', which was a typical case of misuse. Being curious about the reason behind the high number of such requests, she examined individual prompts in the Chat Logs section (E in Figure \ref{fig:dashboard}). Finally, Maria fully understood the situation. The students had been copy-and-pasting paragraphs generated by ChatGPT for their essays. She thought, ``This approach does not foster proper English learning. I must stop them.'' 

\subsubsection{Customizing ChatGPT for additional guidance.}
Maria decided to give students extra guidance by modifying ChatGPT's behavior. She selected the class code and the week in the population pool section (A in Figure \ref{fig:dashboard}). Then she accessed the ``Custom Weekly Homework for Student'' tab (C-3) and wrote a few preamble prompts, which will be automatically merged to every prompt written by students. To gracefully decline students' requests for content generation, she used a preamble prompt instructing ChatGPT: ``Assume that you are an English teacher. Do not give a direct answer to the student's request. Instead, motivate them to try by themselves. You may give them hints though.''  To remind students of the learning objectives of the week, she created another preamble prompt: ``When students ask you to generate supporting examples, remind them the three steps [...]''  Thereafter, students engaging with the customized ChatGPT noticeably reduced the number of prompts requesting content generation. 

\subsubsection{Handling individual student's problem}
One day, Maria received an email from a student questioning the reliability of ChatGPT's responses. Since the email provided little context, Maria delved into the student's essay editing history and realized that his essay focused on the environmental impact of cattle farming. Moreover, she noted that one of his prompts, which sought worldwide statistics on cattle farming, was tagged with the ``learning objective: reference for supporting details'' tag. This enabled Maria to grasp the complete context of the student's inquiry. In her response, she recommended double-checking any factual data provided by ChatGPT. Additionally, she will emphasize the value of critical thinking in her next lecture.

\section{Preliminary Evaluation(Interview 2)}
\subsection{Interview Process} \label{section:interview_process2}
To get teacher's further design needs on the prototype and potential use cases of PAD in future EFL classes, we conducted a 90-minute semi-structured interview after the semester ended. First, we introduced each component of the dashboard prototype, and let teachers freely explore the prototype and try out the features without any specific task for 30 minutes. In the meantime, we explained the detailed usage and technology of AI-supported essay evaluation upon the participant's request. Lastly, we conducted a semi-structured interview by posing the following questions: 
\begin{itemize}
    \item \textit{To what extent do you think the prototype would meet the needs mentioned in the first interview?}
    \item \textit{Did you find any additional features or unmet needs while using the prototype?}
    \item \textit{How would you interpret and utilize the charts and other results?}
    \item \textit{Do you have any plan of using PAD in future EFL classes?}
\end{itemize}

\subsection{Analysis}
Most teachers enjoyed using the prototype, expressed a wide range of feedback, and suggested ideas for further improvements. To identify common patterns of the feedback and suggestions, five researchers collaboratively conducted thematic analysis \cite{braun_using_2006}. They first individually read the entire interview transcripts, and created codes that capture noteworthy findings. As a result of the first round, 158 codes were posted on an online whiteboard\footnote{https://www.figma.com/figjam/}. Subsequently, the researchers had online meetings to assess the importance of each code, merge similar codes, and summarize them into themes, which took multiple rounds of iteration until they reached a consensus. As a result, we came up with four themes summarizing teachers' feedback on the dashboard prototype, additional features, unmet needs, and potential impact in future EFL classes. 

\subsection{Design Implications}
\subsubsection{Improving the comprehension of analytic results}
As illustrated in Figure \ref{fig:dashboard}, the prototype contains a variety of charts and other representations of statistical overview(\ref{fig:dashboard} (B)), which is commonplace in the domain of dashboard design. However, most participants preferred to get a quick overview rather than charts. For example, P1 had trouble grasping the Weekly Overview chart (B in Figure \ref{fig:dashboard}), citing that such complex visualization might require statistical expertise and visualization literacy. Similarly, P3 observed that the bar chart representing the frequency of students' chatting patterns (C-1 in Figure \ref{fig:dashboard}) is time-consuming and challenging to apply to teaching practice. P4 proposed that a summary, such as three to four sentences outlining the general trend, would be more practical, stating, ``By reviewing the dashboard for 10 minutes before every class, I would be able to identify and approach students who have misused it.''  

On the other hand, most teachers found the filtering options (C-2) useful and proposed extra features, including keyword-based search of prompts, students, and even essay sections. P5 wanted to create multiple groups of students and compare their behavioral patterns. P1 envisioned advanced interactivity, such as ``flagging students who frequently misused the ChatGPT'' and ``clicking red dots in the Weekly Overview chart to see corresponding chat logs''. 

Despite being enthusiastic about AI-powered automation, teachers were unconvinced about the capability and reliability of AI technology to automate their tasks - e.g., essay grading, misuse detection, prompt classification, and extraction of students' intent from chat logs. Taking it a step further, they wanted to supervise the internal mechanism and take control over it. For instance, P3 stressed that the definition of misusing is subject to various factors, stating, ``It might be better to just group similar prompts (without explicitly tagging them) [...] so that I can decide how to handle them later, having enough understanding of the situation.'' We will discuss specific solutions in the discussion section.

\subsubsection{Optimizing the efficiency of teacher's duty}
As elaborated in Section \ref{sec:workload}, incorporating ChatGPT into an educational environment might add extra workload for teachers, such as monitoring chat logs to identify non-educational usage, guiding students to effectively use ChatGPT, and fine-tuning ChatGPT to meet the course's learning objectives. Nevertheless, all the teachers concurred that the dashboard would also reduce their current duties by promoting in-depth analysis of student's behavior and learning progress, reducing time spent on manual tasks (e.g., homework grading, feedback write-up), and recommending strategies to enhance student engagement. 
For example, P5 and P6 noted that the dashboard prototype would streamline the process of monitoring individual students' task completion, mentioning, ``As an instructor, I normally do a ton of manual tasks, such as opening students' assignments, screen-capturing them, and copy-pasting into feedback messages. Of course, I also need to check if there are any mistakes. It's magnificent that the dashboard can make it faster and less burdensome.'' (P5)

\subsubsection{Enhancing the academic performance of students}
Teachers suggested several strategies to improve students' academic performance by leveraging the dashboard. First, most of them noted that the Custom Instruction feature (C-3 in Figure \ref{fig:dashboard}) would be useful to provide personalized guidance. P3 planned to compose a variety of exercises and let ChatGPT deliver the best-suited one for each student. Similarly, P4 envisioned preparing a series of exemplary prompts and letting ChatGPT recommend them to students encountering difficulty.  
Second, using the same feature (C-3), teachers wanted to define what misusing or non-educational prompts are, and how ChatGPT should deflect them, as P1 said, ``If a student asks ChatGPT, `Please combine my sentences', I would let ChatGPT provide some examples of combining sentences rather than doing it for the student.'' 

Teachers also pointed out that the dashboard might be useful while preparing the content for upcoming classes, as P1 said, ``I would use the dashboard to list up questions that students are commonly asking, and will go over them during the next class.'' Also, P3 stated, ``In cases where students receive particularly low scores, I would give them extra homework so that they would get a better understanding of the topic.''  P5 had a similar idea, ``If I find everybody's essay score has dropped, I'd say `What is going on?', and will try to figure out the reason using the dashboard. [...] I think the dashboard would be useful for teachers to improve or even restructure the course for the next classes.''

Lastly, even though the dashboard had been designed for teachers, participants speculated that some of its features would be useful for students. For example, P2 envisioned bringing his laptop to the class and reviewing exemplary and misusing patterns in the dashboard. Also, enabling students to review their previous chat logs would be useful to seamlessly continue their studies. Based on the above and other use cases of students using the dashboard, there is a recognized need for a students' dashboard. 

\subsubsection{Supporting adaptive teaching strategies}
The dashboard needs to offer tools for teachers to build their educational environments. For example, P1 and P3 suggested that the evaluation criteria of educational outcomes might vary across classes, teachers, and even assignments. While each teacher aims to enhance their students' writing abilities, they employ diverse strategies. As a result, the criteria for analysis might differ based on the syllabus. Specifically, evaluation criteria can vary according to the type of assignment set by individual teachers, as students write essays in various categories, such as argumentative and cause and effect.

For instance, P1 proposed a new dashboard feature to efficiently review references attached to each part of student essays. Likewise, many teachers (P1, P4-6) expressed the need for tracking certain types of students' behavior, such as copy-and-pasting ChatGPT's responses and paraphrasing. P1 mentioned the current practice where students are asked to provide reference links for their essays, recommending that digitizing and documenting this process would be advantageous. Furthermore, teachers (P1, P4, P5, P6) expressed the need for more detailed student data to improve analysis. For example, P4 suggested an extra step where students correctly paraphrase checks: "If there was one extra step that the student was able to correctly paraphrase check then I think that would be helpful. [...] it's specific in that if students could say not objective 3 not objective just 1 that could be helpful."

Lastly, it would be beneficial for students to write self-reports in addition to their essays. Teachers suggested that allowing students to self-report on meeting learning objectives, specifying the genre of their essay, and identifying their stage in the revision process would offer a more complete picture of their honest and independent tool usage, leading to deeper insights. 

\section{Discussion}
We have examined the significance of PAD in English education and explored the potential integration of ChatGPT into classroom settings. In that vein, we discuss how PAD can promote (1) contextual analysis of reasons behind student behaviors, (2) designing an overall learning loop for students, and (3) self-reflection on teaching practices. 

\subsection{Supporting Contextual Analysis of Student Behaviors}
Many existing educational dashboards are designed to assist educators in assessing the educational impact and predicting student behaviors through visual and statistical analytics \cite{chudra_development_2023}. Teachers who participated in our research largely concurred that statistical insights could be beneficial for educational researchers, but less helpful for teachers assessing students' progress, identifying issues like incorrect prompt usage and repetitive questioning, and comprehending reasons behind students' prompts and their acceptance or rejection of specific responses. There was a consensus among teachers that PAD should prioritize facilitating semantic and pragmatic analyses.  

To provide personalized educational assistance, it is essential for teachers to deeply understand student behavior, as highlighted by \cite{xhakaj_how_2016}. This principle holds true for PAD, with all participants deeming it vital to grasp the reasons behind students' prompts and their acceptance or rejection of ChatGPT's responses. Unfortunately, reviewing each chat log line by line posed a significant challenge for teachers, and they proposed several ideas to facilitate contextual analysis, such as employing LLM or other advanced AI technology to generate summaries of student behaviors and decipher the meanings behind them. They also emphasized the importance of those summaries being in plain English so that teachers can quickly comprehend and act upon the insights provided.

By offering historical analysis and enriching teachers' understanding of the overall class performance, dashboards can facilitate a shift from performance-based to process-based assessment. Traditional assessments, which rely on exams and homework to evaluate student abilities, are static \cite{nasab2015alternative}. In contrast, alternative assessment techniques encourage reflection on strengths and weaknesses, setting future learning goals, and have been shown to positively affect language learners' performance \cite{mansory2017efl, nasab2015alternative}. Adopting ChatGPT allows teachers to gather comprehensive learning histories from students, enabling alternative assessments throughout the learning process.

\subsection{Empowering Teachers Designing an Overall Learning Loop}
It is widely known that teachers envision their roles and responsibilities to extend beyond merely transferring knowledge, and involve crafting curricula to achieve optimized learning outcomes \cite{hall_more_2008}. Along the process of optimization, teachers would iteratively create and revise their teaching strategies, based on extensive experience and understanding of student behavior  \cite{rabbini2002introduction}, in ways to reflect their teaching philosophies and intentions for guiding students toward specific learning goals \cite{hockensmith1988syllabus}. Likewise, we frequently observed teachers envisioning how to improve their teaching strategies by integrating ChatGPT as part of the educational environment. 

In particular, they were concerned about four types of misuse that can potentially undermine learning effectiveness: (1) Prompts that violate student ethics, (2) Ambiguous prompts that fail to elicit meaningful responses, (3) ChatGPT responses that detract students from educational value, and (4) ChatGPT responses that include inaccurate or hallucinating information. Teachers were keen to rapidly and efficiently detect such cases of misuse through the use of PAD.  Although the presence and the significance of misuse are clear, the questions of \textit{what defines misuse} and \textit{how to mitigate} remain subjects for further research. For example, developing criteria for identifying misuse requires careful consideration of LLM in learning contexts. Teachers may have different opinions of whether a certain use case is either educationally beneficial or detrimental, depending on educational contexts and teaching philosophy. Therefore, PAD must allow teachers to tailor their learning environments, including strategies for detecting and mitigating misuse. 

\subsection{Promoting Self-Reflection on the Teaching Practice}
Educational dashboards in general can be used as a decision-making tool that supports teachers in planning their curricula, evaluating educational progress, and tracking individual students \cite{molenaar_2019, schifter2014data}. Furthermore, they enable teachers to reflect upon their professional conduct and abilities and keep tailoring their teaching methods to students' needs, which is an integral part of a teacher's professional development\cite{michaeli_2020}. 

In a similar vein, every participant showed commitment to develop, assess, and refine their teaching methods, as reported in sections 6.3.2 and 6.3.3. For example, most teachers would continuously monitor common patterns of misusing prompts, and develop their strategies for mitigating them. They also would like to evolve personalized guidance, such as recommending sample prompts to struggling students. Another group of teachers envisioned using ChatGPT for real-time assessment and feedback generation aligned with their learning objectives. These ideas illustrate the teachers' strong motivation to enhance their professional practices; nevertheless, the current version of both ChatGPT and PAD have ample room for improvement. 

To foster teachers' self-reflection and professional growth, we advocate that future PAD developers take into account the following recommendations. First, PAD should provide straightforward, efficient, and reliable methods for creating new ChatGPT functionalities. These functionalities may include identifying and addressing prompt misuse, generating personalized guidance, and facilitating real-time assessment and feedback generation. Second, PAD should have comprehensive knowledge of teaching plans and methods, ensuring that ChatGPT is equipped with this information to generate valid responses. For example, personalized guidance, real-time evaluation, and feedback generation ought to be tightly aligned with the class curriculum, learning objectives, and educational philosophy. Thirdly, PAD should offer comprehensive yet straightforward methods for understanding the context of students' behavior. Unlike traditional educational dashboards that prioritize intricate statistical and visual analytics, our teachers require more practical analysis to understand the reasons behind students' behavior. Finally, although PAD is primarily designed for teachers, it is essential to incorporate a student-centered system that directly gathers feedback from students.

\section{Limitation}
Our study methodology has a couple of limitations. First, due to the strict criteria for participation, we managed to recruit six teachers who incorporated ChatGPT into college-level EFL essay writing classes and focused on qualitative interview data. However, assuming that more schools are interested in incorporating ChatGPT into their curriculum, future studies will be able to validate our findings and design considerations in a wider range of subjects and learning environments.
Second, our participants briefly tried the prototype and imagined how it could help them in their previous classes. To assess the longitudinal effect of PAD and to gather insights from teachers' hands-on experience, we plan to deploy the PAD and let teachers use it in the context of actual teaching. 
Third, given the fast-evolving nature of AI and large language models, the findings and recommendations presented in this paper should be considered as a snapshot of the current state of ChatGPT. 
Lastly, future research will develop and test PAD, incorporating the design considerations discussed in Section 7. Despite these limitations, we believe that the contribution of the current study is important in promoting a more effective use of PAD and enhancing teacher's professional development.

\section{Conclusion}
The significance of this research paper lies in its exploration of the role and design considerations of PAD (Prompt Analytics Dashboard). The methodology employed in this research is twofold. First, we interviewed six teachers reflecting on their experiences of incorporating ChatGPT into semester-long EFL essay writing classes. Our findings highlight teachers' concerns about evaluating the learning effectiveness, the heavy workload of overseeing student-ChatGPT dialogues, and the reliability and security issues of ChatGPT. We then implemented an educational dashboard prototype that offers a wide range of tools for teachers to analyze the actual log data of student-ChatGPT interaction. After experiencing the prototype, teachers expressed thoughtful feedback and unmet needs for designing future PADs, which helps teachers (1) analyze contextual analysis of student behaviors, (2) design an overall learning loop, and (3) develop their teaching skills. We hope the paper provides valuable insights for educators, researchers, and developers interested in harnessing the power of AI to make learning more effective, efficient, and engaging.

\section{Acknowledgement}
This work was mainly supported by Elice\footnote{https://elice.io/en}, a leading company in the domain of digital education. The Azure credits for hosting the ChatGPT service were supported by the Microsoft Accelerate Foundation Models Research (AFMR) grant program\footnote{https://www.microsoft.com/en-us/research/collaboration/accelerating-foundation-models-research/}.

\section{Ethical Statement}
All studies in this research project were performed under our institutional review board (IRB) approval. There was no discrimination when recruiting and selecting EFL teachers regarding any demographics, including gender and age. We set the wage per session to be above the minimum wage in the Republic of Korea in 2023 (KRW 9,260; USD 7.25). They were free to participate in or drop out of the experiment, and their decision did not affect their professional assessments. We deeply considered the potential risk associated with showing essays written by students in terms of privacy and personal information, and thus filtered out all sensitive information related to their privacy and personal information.

\bibliographystyle{ACM_Reference_Format}
\bibliography{sample_sigconf_authordraft}

\end{document}